\documentclass[11pt]{llncs}

\usepackage[T1]{fontenc}
\usepackage{ae,aecompl}
\usepackage{amsmath}
\usepackage{amssymb}
\usepackage{algorithm}
\usepackage{algorithmic}
\usepackage{graphicx}
\usepackage{wrapfig}
\usepackage{bbm}
\usepackage[bbgreekl]{mathbbol}
\graphicspath{{Figures/}}

\newcommand{\anc}{\ensuremath{\mathrm{wla}}}
\newcommand{\wla}{\ensuremath{\mathrm{wla}}}
\newcommand{\rrank}{\ensuremath{\mathrm{rank}}}
\newcommand{\occ}{\ensuremath{\mathrm{occ}}}
\newcommand{\ndocs}{\ensuremath{\mathrm{ndocs}}}
\newcommand{\Lookup}{\ensuremath{Lookup}}

\newcommand{\oT}{\ensuremath{\overline{\mathcal{T}}}}
\newcommand{\oF}{\ensuremath{\overline{{\mathcal{F}}}}}
\newcommand{\tT}{\ensuremath{\widetilde{\mathcal{T}}}}
\newcommand{\bT}{\ensuremath{\mathbbm{T}}}
\newcommand{\cT}{\ensuremath{{\cal T}}}
\newcommand{\bv}{\ensuremath{\mathbbm{v}}}
\newcommand{\bu}{\ensuremath{\mathbbm{u}}}
\newcommand{\bp}{\ensuremath{\mathbbm{p}}}
\newcommand{\cF}{\ensuremath{\mathcal{F}}}
\newcommand{\caT}{\ensuremath{{\cal T}}}

\newcommand{\no}[1]{}

\makeindex

\begin{document}

\title{Cross-Document Pattern Matching}

\institute{\empty}

\author
{
Gregory Kucherov\inst{1}
\and
Yakov Nekrich\inst{2}
\and
Tatiana Starikovskaya\inst{3,1}
}
\institute{
Laboratoire d'Informatique Gaspard Monge, Universit\'e Paris-Est \& CNRS, Marne-la-Vall\'ee, Paris, France, \email{Gregory.Kucherov@univ-mlv.fr}
\and
Department of Computer Science, University of Chile, \email{yakov.nekrich@googlemail.com}
\and
Lomonosov Moscow State University, Moscow, Russia, \email{tat.starikovskaya@gmail.com}
}

\date{\empty}

\maketitle

\begin{abstract}
We study a new variant of the string matching problem called {\em
  cross-document string matching}, which is the problem of indexing a
collection of documents to support an efficient search for a pattern in
a selected document, where the pattern itself is a substring of
another document. Several variants of this problem are considered, and
efficient linear-space solutions are proposed with query
time bounds that either do not depend at all on the pattern size or
depend on it in a very limited way (doubly logarithmic). As a side result,
we propose an improved solution to the {\em weighted level
  ancestor} problem. 
\end{abstract}

\section{Introduction}
\no{Identifying occurrences of a pattern $P$ in a text $T$ is one of the most 
fundamental and most extensively studied problems in string algorithms. It is 
well known that we can construct a linear space data structure that 
finds all occurrences of a query substring $P$   in optimal $O(|P|+\occ)$ time, 
where 
$|P|$ denotes the length of $P$  and $\occ$ is the number of occurrences
 of $P$ in $T$.}
In this paper we study the following variant of the string matching
problem that we call {\em cross-document string matching}: given a collection of strings (documents) stored in a
``database'', we want to be able to efficiently search for a pattern in
a given document, where the pattern itself is a substring of another
document. More formally, assuming we have a set of documents
$T_1,\ldots,T_m$, we want to answer queries about the
occurrences of a substring $T_k[i..j]$ in a document $T_\ell$. 

This scenario may occur in various practical situations when we have
to search for a pattern in a text stored in a database, and the
pattern is itself drawn from a string from the same database. This is
a common situation in bioinformatics, where one may want to repeatedly
look for
genomic elements drawn from a genome within a set of genomic
sequences involved in the project. In bibliographic
search, it is common to look up words or citations
coming from one document in other documents. Similar scenarios may
occur in other traditional applications of string matching, such as in
the analysis of web server logfiles for example. 


In this paper, we study different versions of the cross-document
string matching problem. First, we distinguish between counting and
reporting queries, 
asking respectively about the number of occurrences of $T_k[i..j]$ in
$T_\ell$ or about the occurrences themselves. The two query types
lead to slightly different solutions. In particular, the counting
problem uses the {\em weighted level ancestor} problem
\cite{Farach:1996:PHS:647815.738452,Amir:2007:DTS:1240233.1240242} to 
which we propose a new solution with an improved complexity bound. 

We further consider different variants of the two problems. The first
one is the dynamic version where new documents can be 
added to the database. In another variant, called {\em document counting
  and reporting}, we only need to respectively count or report the
documents containing the pattern, rather than counting or reporting
pattern occurrences within a given document. This version is very close to the
{\em document retrieval problem} previously studied 
(see \cite{DBLP:conf/soda/Muthukrishnan02} and
later papers referring to it), with the difference that in our case the pattern is
itself selected from the documents stored in the database. Finally, we
also consider \emph{succinct} data structures for the above problems, where 
we  keep involved index data structure in compressed form.

Let $m$ be the number of stored strings and $n$ the total length of
all strings. Our results are summarized below:
\begin{itemize}
\item[(i)] for the counting problem, we propose a solution with query time
  $O(t+\log\log m)$, where $t=\min(\sqrt{\log \occ/\log \log \occ}, \log \log 
|P|)$,  $P=T_k[i..j]$ is the searched substring and $\occ$ is the number of its occurrences in $T_\ell$,
\item[(ii)] for the reporting problem, our solution outputs all the
  occurrences in time $O(\log\log m+\occ)$, 
\item[(iii)] in the dynamic case, when new documents can be dynamically added
  to the database, we are able to answer counting queries in time
$O(\log n)$
and reporting queries in time $O(\log
  n+\occ)$, whereas the updates take time $O(\log n)$ per character,
\item[(iv)] for the document counting and document reporting problems, our
  algorithms run in time $O(\log n)$ and $O(t+\ndocs)$ respectively, where
  $t$ is as above and $\ndocs$ is the number of reported documents,
\item[(v)] finally, we also present  succinct data structures that support 
 counting, reporting, and document reporting queries in cross-document scenario (see Theorems~\ref{theor:compact1} and~\ref{theor:compact2} in Section~\ref{sec:compact}).
\end{itemize}
For problems (i)-(iv), the involved data structures occupy $O(n)$
space. 
Interestingly, in the cross-document scenario, the query
times either do not depend at all on the pattern length or
depend on it in a very limited (doubly logarithmic) way. 

Throughout the paper 
positions in strings are numbered from $1$. Notation $T[i..j]$ stands
for the subword $T[i]T[i+1]\ldots T[j]$ of $T$, and $T[i..]$ denotes
the suffix of $T$ starting at position $i$.  



\section{Preliminaries}

\subsection{Basic Data Stuctures}
	\label{sec:prelim}


We assume a basic knowledge of suffix trees and suffix arrays. 

Besides using suffix trees for individual strings $T_i$, we will 
also be using the {\em generalized suffix tree} for a set of strings
$T_1, T_2, \ldots, T_m$ that can be viewed as the suffix tree for the
string $T_1\$_1 T_2\$_2 \ldots T_m\$_m$. A leaf in a suffix tree for $T_i$ is 
associated with a distinct suffix of $T_i$, and a leaf in the
generalized suffix tree is associated with a suffix of some document
$T_i$ together with the index $i$ of this document. 
%
%
%
%
We assume that for each node~$v$ of a suffix tree, the number $n_v$
of leaves in the subtree rooted at $v$, as well as its string
depth~$d(v)$ can be recovered in constant time. Recall that the string
depth $d(v)$ is the total length of strings labelling the edges along
the path from the root to $v$. 



We will also use the suffix arrays for individual documents as well as
the {\em generalized suffix array} for strings $T_1, 
T_2, \ldots, T_m$. 
Each entry of the suffix array for $T_i$ is associated with a
distinct suffix of $T_i$ and each entry of the generalized suffix
array for $T_1,\ldots, T_m$ is associated with a suffix of some
document $T_i$ and the index $i$ of the document the suffix comes
from. We store these document indices in a separate array $D$, called
{\em document array}, such that $D[i]=k$ if the  $i$-th entry of the
generalized suffix array for $T_1,\ldots, T_m$ corresponds to a
suffix coming from $T_k$. 


For each considered suffix array, we assume available, when needed, two auxiliary
arrays: an inverted suffix array and another array, called the
LCP-array, of longest common prefixes between each suffix and the preceding one in the lexicographic order. 

Suffix trees and suffix arrays are naturally related: if the children
of any internal node of a suffix tree are ordered in the lexicographic
order of the labels (actually, of the first symbols of the labels, as they are
all distinct), then the leaves ordered ``left-to-right'' correspond
exactly to the suffix array with respect to the referred
suffixes. 
\no{
The suffix tree will be said to be {\em ordered} when we
explicitly require it to be ordered as above. In this case, we will
freely navigate between suffix tree leaves and suffix array entries
which follow the same order. 
}




\subsection{Weighted Level Ancestor Problem}
	\label{sec:wla}
	The {\em weighted level ancestor} problem, defined in~\cite{Farach:1996:PHS:647815.738452}, is a generalization of the level ancestor
 problem~\cite{BerkmanV94,BenderF04}
for the case when tree edges are assigned positive weights.


Consider a rooted tree $\cT$ whose edges are assigned positive
integer weights. For a node $w$, let $weight(w)$ denote the total weight of
the edges on the path from the root to $w$. $depth(w)$ denotes the
usual tree depth of $w$. 
A {weighted level ancestor} query $\anc(v,q)$ asks, given a node $v$ and
a positive integer $q$, for the ancestor $w$ of $v$ of minimal depth such that
$weight(w)\ge q$ ($\anc(v,q)$ is undefined if 
there is no
such node $w$). 


Two previously known
solutions~\cite{Farach:1996:PHS:647815.738452,Amir:2007:DTS:1240233.1240242}
for weighted level ancestors problem
achieve $O(\log \log W)$ query time using linear space, where $W$ is the 
total weight of all tree edges.
Our data structure also uses $O(n)$ space, but achieves a faster query time in 
many special cases.
We prove the following result.
\begin{theorem}\label{theor:wla}
There exists an $O(n)$ space data structure that answers 
weighted ancestor query $\anc(v,q)$ in  
$O(\min(\sqrt{\log g/\log \log g}, \log \log q))$ time, 
where $g=\min(depth(\anc(v,q)),depth(v)-depth(\anc(v,q)))$. 
\end{theorem}

If every internal node is a branching node, we obtain the following 
corollary.

\begin{corollary}\label{cor:wla}
Suppose that every internal node in $\cT$ has at least two children. There exists 
an $O(n)$ space data structure that finds $w=\anc(v,q)$ in 
$O(\sqrt{\log n_w/\log \log n_w})$ time, where $n_w$ is the number of 
leaves in the subtree of $w$. 
\end{corollary}


Our approach combines the heavy path decomposition technique
  of~\cite{Amir:2007:DTS:1240233.1240242} 
with efficient data structures for finger searching in a set of
  integers. Due to space limitations, the proof is given in the
  Appendix.

\section{Cross-document Pattern Counting and Reporting}
\label{sec:cross-doc}

\subsection{Counting}
	\label{sec:counting}
	In this section we consider the problem of counting occurrences of 
a pattern $T_k[i..j]$ in a document $T_\ell$. 

\no{
Assume we are given a set of documents $T_1,T_2,\ldots,T_m$. We first
consider the problem of answering counting queries: how many
occurrences of $T_k[i..j]$ are there in $T_\ell$? A query submits then
four parameters: $k$, $i,j$ and $\ell$. 
}

Our data structure consists of the generalized suffix array $GSA$ for 
documents $T_1,\ldots, T_m$ and individual suffix trees $\caT_i$ 
for every document $T_i$.
We assume that entries of $GSA$ and 
leaves of suffix trees $\caT_i$ are linked by pointers so that given the 
location of some suffix $T_k[i..]$ in $GSA$, 
we can retrieve its position in $\caT_k$.  

\no{
All documents $T_1,\ldots, T_m$ are stored in a generalized suffix array $GSA$.
We augment $GSA$ with the document array $D$: $D[r]=l$ if  $GSA[r]$ 
is a suffix of $T_{\ell}$. 
Besides that, we also construct a suffix tree $\caT_i$ for each document 
$T_i$. 
 We assume that instances of the same suffix in the array $GSA$ and 
suffix tree $\caT_i$ are connected by pointers. Thus if we know the 
position of some suffix $T_k[i..]$ in the generalized suffix array, 
we can identify its position in $\caT_k$ and vice versa.  
}

For every suffix tree $\caT_{\ell}$ we store a data structure of
 Theorem~\ref{theor:wla} supporting weighted level ancestor queries on
 $\caT_{\ell}$. 
We also augment the document array $D$ with a $O(n)$-space data structure that 
answers queries $rank(k,i)$ (number of entries storing $k$ before
 position $i$ in $D$)
and  $select(k,i)$ ($i$-th entry from
the left storing $k$). Using the result 
of~\cite{DBLP:conf/soda/GolynskiMR06},  we can support 
such rank and select queries in $O(\log\log m)$ and $O(1)$ time respectively. 
Moreover, we construct a data structure that answers range minima 
queries (RMQ) on the $LCP$ array: 
for any $1\le r_1\le r_2\le n$, find the minimum among 
$LCP[r_1],\ldots LCP[r_2]$. There exists a linear 
space RMQ data structure that supports queries in constant time, see 
 e.g.,~\cite{Bender:2000:LPR:646388.690192}. An RMQ query
on the $LCP$ array computes the length of the longest
common prefix of 
two suffixes $GSA[r_1]$ and $GSA[r_2]$, denoted $LCP(r_1,r_2)$.

\no{
We store the documents $T_1,T_2,\ldots,T_m$ in a generalized suffix
array. We also build an individual suffix tree for each document $T_i$
and link every entry of the generalized suffix array to the
corresponding leaf of the corresponding suffix tree. More
specifically, if an entry corresponds to a suffix $T_k[i..]$ of $T_k$,
then it is linked to the leaf $T_k[i..]$ of the individual sufffix
tree for $T_k$. 

We supplement the generalized suffix array by a {\em document array}
$D$.
On the array $D$, we
build a rank/select data structure as described
in \cite{DBLP:conf/soda/GolynskiMR06}. The data structure takes as
much space as the array $D$ itself and allows answering queries
$rank(k,i)$ (number of entries storing $k$ before position $i$ in $D$)
in $O(\log\log m)$ time, and queries $select(k,i)$ ($i$-th entry from
left storing $k$). It can be constructed in $O(n)$ time. 
}

Our counting algorithm consists of two stages. First, using 
$GSA$, we identify a position $p$ of $T_\ell$ at which 
the query pattern $T_k[i..j]$ occurs, or determine that no such $p$ exists.
Then we find the locus of $T_k[i..j]$ in the 
suffix tree $\caT_{\ell}$ using a weighted ancestor query.

Let $r$ be the position of $T_k[i..]$ in the $GSA$. We find 
indexes $r_1=select(\ell,rank(r,\ell))$ and $r_2=select(\ell,rank(r,\ell)+1)$ 
in $O(\log \log m)$ time. $GSA[r_1]$ (resp. $GSA[r_2]$) is the closest
suffix from document $T_{\ell}$ that precedes (resp. follows) $T_k[i..]$ 
in the lexicographic order of suffixes. 
Observe now that $T_k[i..j]$ occurs in $T_\ell$ if and only if
either $LCP(r_1, r)$ or $LCP(r, r_2)$ (or both) is no less than
$j-i+1$. If this holds, then the starting position $p$ of $GSA[r_1]$ (respectively, starting position of $GSA[r_2]$) is the
position of $T_k[i..j]$ in $T_\ell$. Once such a position $p$ is
found, we jump to the corresponding leaf $T_\ell[p..]$ in  the suffix
tree of $T_\ell$. 

\no{
We now describe our algorithm to count the number of occurrences of
$T_k[i..j]$ in $T_\ell$. 
We first retrieve the rank $r$ of $T_k[i..]$ in 
the generalized suffix array for
$T_1, T_2, \ldots, T_m$. The rank/select
data structure allows us to compute values $r_1 = \min\{t > r | D[t]
= \ell\}$ and $r_2 = \max\{t < r | D[t] = \ell\}$ in
$O(\log{\log{m}})$ time.  

Let $LCP(r',r'')$ be the length of the longest common prefix of
suffixes with ranks $r'$ and $r''$ in the lexicographic order on the
set of suffixes of $T_1, T_2, \ldots, T_m$. Using an RMQ data
structure~\cite{Bender:2000:LPR:646388.690192} on the LCP-array,
$LCP(r',r'')$ can be computed in $O(1)$ time for any $r'$ and $r''$.  
Observe now that $T_k[i..j]$ occurs in $T_\ell$ if and only if
either $LCP(r_1, r)$ or $LCP(r_2, r)$ (or both) is no less than
$j-i+1$. If this holds, then $p=GSA[r_1]$ (respectively $p=GSA[r_2]$) is the
position of $T_k[i..j]$ in $T_\ell$. Once such a position $p$ is
found, we jump to the corresponding leaf $T_\ell[p..]$ in  the suffix
tree of $T_\ell$. 
}

Let $v$ be the leaf of $\caT_\ell$ that contains the suffix $T_\ell[p..]$. 
Then the weighted level ancestor $u=\wla(v,(j-i+1))$ is the locus 
of $T_k[i..j]$ in $\caT_{\ell}$. This is because 
$T_\ell[p..p+j-i]=T_k[i..j]$.
By Corollary~\ref{cor:wla}, we can find node $u$ in $O(\sqrt{\log n_u/\log\log n_u})$ time, where $n_u$ is the number of leaf descendants of $u$. 
Since $u$ is the locus node of $T_k[i..j]$, $n_u$  is the 
number of occurrences of $T_k[i..j]$ in $T_\ell$. 
By Theorem~\ref{theor:wla}, we can find $u$ in $O(\log \log (j-i+1))$ time.

\no{ 
\begin{lemma}
\label{ancestor}
Let $u$ be the highest ancestor of $v$ with string depth at least
$j-i+1$ and $n(u)$ be the number of leaves in the subtree rooted at
$v$. Then the number of occurrences of $T_k[i..j]$ in $T_\ell$ is
equal to $n(u)$. 
\end{lemma}
\begin{proof}
Indeed, any suffix of $T_\ell$ starting with $T_k[i..j]$ ends in a
leaf of the subtree rooted at $v$, and vica versa. 
\end{proof}

To compute the node $u$ from Lemma~\ref{ancestor}, we apply the
weighted level ancestor construction from Section~\ref{sec:wla}.
Let $t=\min\{\log{\log{n}}, \log \log d(u), \sqrt{ \log
q/\log \log q} \}$, where $n$ is the size of the suffix tree, $d(u)$
is the string depth of $u$ and $q$ is the minimum of the length of the
path from the root to $u$ and $n(u)$. \marginpar{\em $n(u)$ is a
number!?} According to Section~\ref{sec:wla}, the node $u$ can be
computed in $O(t)$ time. In conclusion, we obtain the following
theorem.  
}
Summing up, we obtain the following Theorem.
\begin{theorem}
For any $1\le k,\ell \le m$ and $1\le i\le j \le |T_k|$, we can 
count the number of occurrences of $T_k[i..j]$ in $T_\ell$ 
in $O(t+\log\log m)$ time, where
$t=\min(\sqrt{\log \occ/\log \log \occ}, \log \log (j-i+1))$ and 
 $\occ$ is the number of occurrences.
The underlying
indexing structure takes $O(n)$ space and can be constructed in $O(n)$
time. 
\end{theorem}
Observe that our data structure always answers range counting queries 
in $O(\log \log n)$ time. If $m$ and either the
 pattern length $(j-i+1)$ or the number of occurrences
are sufficiently small, the query time is even better. For instance
if $m=O(1)$ and $\occ=O(1)$, a query is answered in constant time. 


\subsection{Reporting}
	\label{sec:reporting}
	A reporting query asks for all  occurrences of a
substring $T_k[i..j]$ in $T_\ell$. 

Compared to counting queries, we make a slight change in the data
structures: instead of using  suffix trees for individual
documents $T_i$, we use  suffix arrays. Similarly to the
previous section, we link each entry of $GSA$ 
to a corresponding entry in the corresponding individual suffix
array. The rest of the data structures is unchanged. 

We first find an occurrence of $T_k[i..j]$ in $T_\ell$ (if there is one) 
with the method described in Section~\ref{sec:counting}. Let
$p$ be the position of this occurrence in $T_\ell$. We then jump to
the corresponding entry $r$ of the suffix array $SA_\ell$ for the document
 $T_\ell$. 
%
%
%
Let $LCP_\ell$ be the LCP-array of $SA_\ell$. 
Starting with entry $r$, we visit adjacent entries $t$ of $SA_\ell$
moving both to the left and to the right as long as $LCP_\ell[t] \geq j-i+1$. 
While 
this holds, we
report $SA_\ell[t]$ as an occurrence of $T_k[i..j]$. 
It is easy to observe that the procedure is correct and that no
occurrence is missing. 
As a result, we obtain the following theorem. 

\begin{theorem}
\label{th:reporting}
All the occurrences of $T_k[i..j]$ in $T_\ell$ can be reported in
$O(\log{\log{m}}+\occ)$ time, where $\occ$ is the number of occurrences. The
underlying indexing structure takes $O(n)$ space and can be
constructed in $O(n)$ time. 
\end{theorem}

Observe that the algorithm has no dependence whatsoever on the pattern
length, and that the query time does not depend on the length of
documents but only on their number.

\section{Variants of the Problem}	

\subsection{Dynamic Counting and Reporting}
	\label{sec:dynamic}
	In this section we focus on a {\em dynamic version} of counting and
reporting problems, where the 
only dynamic operation consists in {\em adding a document to the
database}\footnote{document deletions are also possible to support but
require some additional constructions that are left to the extended
version of this paper}.

Recall that in the static case, counting occurrences of $T_k[i..j]$ in
$T_\ell$ is done through the following two steps
(Section~\ref{sec:counting}):

\begin{enumerate}
\item compute position $p$ of some occurrence of $T_k[i..j]$ in
$T_\ell$, 
\item in the suffix tree of $T_\ell$, find the locus of string 
$T_\ell[p..p+j-i]$, and retrieve
the number of leaves in the subtree rooted at $u$. 
\end{enumerate}
For reporting queries (Section~\ref{sec:reporting}), Step~1 is
basically the same, while Step~2 is different and uses an individual
suffix array for $T_\ell$. 

In the dynamic framework, we follow the same general two-step
scenario. Note first that 
since Step~2, for both counting and reporting, uses data structures for individual documents only, it
trivially applies to the dynamic case without changes. However,
Step~1 requires serious modifications that we describe below. 

Since the suffix array is not well-suited for dynamic updates, 
at Step~1 we will use the generalized suffix tree for
$T_1, T_2, \ldots, T_m$ hereafter denoted $GST$. 
%
For each suffix of $T_1, T_2, \ldots, T_m$ we store a pointer
to the leaf of $GST$ corresponding to this suffix. 
\no{Recall that in the ordered generalized suffix tree the
``left-to-right'' order of leaves corresponds to 
the lexicographic order of suffixes. }
%
Unfortunately, 
it is not easy to maintain the lexicographically ordered list of
suffixes when $GST$ is dynamically updated, as it is not easy to
quickly determine the location of a newly created leaf in the list. 
Another task to be solved is to support updates of LCP-values
and range minima queries on these values\footnote{supporting dynamic
RMQ could be done with the general method
of \cite{Brodal:2011:PMQ:2033190.2033215}, however we will give here a
simpler {\em ad hoc} algorithm with the same time complexity, which is sufficient for our purposes}. 

To this end, we introduce the following additional data
structure. We maintain a dynamic doubly-linked
list corresponding to the Euler tour of the current $GST$. This list is
 denoted by $EL$. Each
internal node
of $GST$ is stored in two copies in $EL$, corresponding
respectively to the first and last visits of the node during the Euler
tour. 
Leaves of $GST$
are kept in one copy. Observe that the leaves of $GST$ appear in $EL$
in the same ``left-to-right'' order, although not consecutively. 

On $EL$, we maintain the data
structure of \cite{Bender:2002:TSA:647912.740822} which allows, given
two list elements, to determine their order in the list in $O(1)$ time
(see also \cite{Dietz:1987:TAM:28395.28434}). Insertions 
of elements in the list are supported in $O(1)$ time too. 

Furthermore, we maintain a balanced  tree, denoted $BT$, whose
leaves are elements of $EL$.  
Note that the size of $EL$ is bounded by $2n$ ($n$ is the size
of $GST$) and the height of $BT$ is $O(\log n)$.
Since the leaves of $GST$ are a subset of the leaves of $BT$, we call
them {\em suffix leaves} to avoid the ambiguity. 

Each internal node $u$ of $BT$ stores two kinds of
information: (i) the rightmost and leftmost suffix leaves in the
subtree of $BT$ rooted at $u$, (ii) minimal LCP value among all
suffix leaves in the subtree of $BT$ rooted at $u$. 





Finally, we will also need an individual suffix array for each inserted
document $T_i$. 

We are now in position to describe the algorithm of Step~1. 
Like in the static case, we first retrieve the leaf of $GST$
corresponding to suffix $T_k[i..]$.
To identify a position of an
occurrence of $T_k[i..j]$ in $T_\ell$, we have to examine 
the two closest elements in the list of leaves of $GST$, one from
right and from left, corresponding to suffixes of $T_\ell$. 
To find these two suffixes, we perform a binary search on the suffix
array for $T_\ell$ using order queries
of \cite{Bender:2002:TSA:647912.740822} on $EL$. This step takes
$O(\log |T_\ell|)$ time. 

We then check if at least one of these two suffixes corresponds to an
occurrence of $T_k[i..j]$ in $T_\ell$. In a similar way to
Section~\ref{sec:cross-doc}, we have to compute the longest common
prefix between each of these two suffixes and $T_k[i..]$, and compare
this value with $(j-i+1)$. This amounts
to computing the minimal $LCP$ value among all the suffixes of the
corresponding range, i.e. to answering a range-minima query. To do
this, we resort to the list $EL$ and the  tree $BT$ and use the
standard technique used for answering  range queries: for any 
sublist $L'$ of $EL$ we can identify $O(\log n)$ nodes $v_i$ of $BT$, so 
that an element $e$ belongs to $L'$ if and only if it is a leaf descendant 
of some node $v_i$. 
We retrieve $O(\log n)$ nodes $v_i$ that cover the relevant sublist of $EL$.
 The least among all minimal $LCP$ values stored in nodes $v_i$ is the 
minimal LCP value for the specified range of suffixes. 
Thus, computing the length of the longest common prefix of two suffixes
 takes $O(\log n)$ 
time.  
\no{
 We retrieve
the corresponding sublist of $EL$ and extract $O(\log n)$ nodes of
$BT$ that cover the sublist. The minimum among
all $LCP$ values stored in these nodes yields the desired value. The
whole computation is done in $O(\log n)$ time. 
}
Once a witness occurrence of $T_k[i..j]$ in $T_\ell$ is found, Step~2
is done as explained in 
Sections~\ref{sec:counting},\ref{sec:reporting}. 

The query time bounds are summarized in the following lemma.
\begin{lemma}
Using the above data structures, counting and reporting all
occurrences of $T_k[i..j]$ in $T_\ell$ can be done respectively in 
time $O(\log n)$ and time 
$O(\log n + \occ)$, where
$\occ$ is the
number of reported occurrences. 
\end{lemma}

We now explain how the involved data structures are updated. Suppose
that we add a new document $T_{m+1}$. Extending the generalized suffix
tree by $T_{m+1}$ is done in time $O(|T_{m+1}|)$ by McCreight's or
Ukkonen's algorithm, i.e. in $O(1)$ amortized time per symbol. 

When a new node $v$ is added to a suffix tree, the following updates
should be done (in order):
\begin{itemize}
\item[(i)] insert $v$ at the right place of the list $EL$ (in two copies
if $v$ is an internal node),
\item[(ii)] rebalance the tree $BT$ if needed,
\item[(iii)] if $v$ is a leaf of $GST$ (i.e. a suffix leaf of $BT$),
update LCP values and rightmost/leftmost suffix leaf information in
$BT$,
\end{itemize}

To see how update (i) works, we have to recall how suffix tree is
updated when a new document is inserted. Two possible updates are
creation of a new internal node $v$ by splitting an edge into two (edge
subdivision) and creating a new leaf $u$ as a child of an existing
node. In the first case, we insert the first copy of $v$ right after
the first copy of its parent, and the second copy right before the
second copy of its parent. In the second case, the parent of $u$ has
already at least one child, and we insert $u$ either right after the
second (or the only) copy of its left sibling, or right before the
first (or the only) copy of its right sibling. 

Rebalancing the tree $BT$ (update (ii)) is done using standard
methods. 
Observe that during the
rebalancing we may have to adjust the 
LCP and rightmost/leftmost suffix leaf information for internal nodes,
but this is easy to do as only a finite number of local modifications
is done at each level. 

Update (iii) is triggered when a new leaf $u$ is created in $GST$ and
added to $EL$. First of all, we have to compute the $LCP$ value for $u$
and possibly to update the $LCP$ value of the next suffix leaf $u'$ to
the right of $u$ in $EL$. This is done in $O(1)$ time as follows. At
the moment when $u$ 
is created, we memorize the string depth of its parent
$D=d(parent(u))$. Recall that 
the parent of $u$ already has at least one child before
$u$ is created. If $u$ is neither the leftmost nor the rightmost child of its
parent, then we set $LCP(u)=D$ and $LCP(u')$ remains unchanged (actually it
also equals $D$). If $u$ is the leftmost child of its parent, then we
set $LCP(u)=LCP(u')$ and then $LCP(u')=D$. Finally, if $u$ is the
rightmost child, then $LCP(u)=D$ and $LCP(u')$ remains
unchanged. 

We then have to follow the path in $BT$ from the new leaf $u$ to the
root and possibly update the LCP and rightmost/leftmost suffix leaf
information for all nodes on this path. These updates are
straightforward. Furthermore, during this traversal we also identify
suffix leaf $u'$ (as the leftmost child of the first right sibling
encountered during the traversal), update its $LCP$ value and, if
necessary, the $LCP$ values on the path from $u'$ to the root of
$BT$. All these steps take time $O(\log n)$.

Thus, updates of all involved data structures take  $O(\log n)$ time per symbol. 
%
The following theorem summarizes the results of this section. 

\begin{theorem}
In the case when documents can be added dynamically, the number of
occurrences of $T_k[i..j]$ in $T_\ell$ 
can be computed in time $O(\log n)$
and reporting these occurrences can be done in time 
$O(\log n + occ)$, where
$occ$ is their
number. The underlying data structure occupies
$O(n)$ space and an update takes $O(\log n)$ time per
character. 
\end{theorem}

\subsection{Document Counting and Reporting}
	\label{sec:doc}
	Consider a static collection of documents $T_1,\ldots,T_m$. In this section
we focus on document reporting and counting queries: report or count
the documents which contain at least one occurrence of $T_k[i..j]$, for some
$1\leq k\leq m$ and $i\le j$. 

For both counting and reporting, we use the generalized suffix
tree, generalized suffix array 
and the document array $D$ for $T_1, T_2, \ldots, T_m$.
We first retrieve the leaf of the
generalized suffix tree labelled by $T_k[i..]$ and compute its highest
ancestor $u$ of string depth at least $j-i+1$, using the weighted level
ancestor technique of Section~\ref{sec:wla}. The
suffixes of $T_1, T_2, \ldots, T_m$ starting with $T_k[i..j]$
(i.e. occurrences of $T_k[i..j]$) correspond then to the leaves of the
subtree rooted at $u$, and vice  versa. 
As shown in section~\ref{sec:counting}, this step takes 
$O(t)$ time, where
$t=\min(\sqrt{\log \occ/\log \log \occ}, \log \log (j-i+1))$ and 
 $\occ$ is the number of occurrences of $T_k[i..j]$ (this time in all
 documents).

Once $u$ has been computed, we retrieve the interval
$[left(u)..right(u)]$ of ranks of all the leaves under interest. We are
then left with the problem of counting/reporting distinct values
in $D[left(u)..right(u)]$. This problem is exactly the same as the 
color counting/ color reporting problem that has been studied extensively
(see e.g.,~\cite{Gagie:2010:CRQ:1928328.1928337} and references therein).



For color reporting queries, we can use the  solution
of \cite{DBLP:conf/soda/Muthukrishnan02} based on a $O(n)$-space
data structure for RMQ, applied to (a transform of)
the document array $D$. The pre-processing time is $O(n)$. Each
document is then reported in $O(1)$ time, 
i.e. all relevant documents are reported in $O(\ndocs)$ time, where 
$\ndocs$ is their number. The whole reporting query then takes time
$O(t+\ndocs)$ for $t$ defined above. 

For counting, we use the solution described in~\cite{conf/icalp/BozanisKMT95}.
 The data structure 
requires $O(n)$ space and a color counting query takes $O(\log{n})$
time. The following theorem presents a summary. 

\begin{theorem}
We can store a collection of documents $T_1,\ldots, T_m$ in a linear 
space data structure, so that for any pattern $P=T_k[i..j]$ 
all documents that contain $P$ can be reported and counted in 
$O(t+\ndocs)$ and $O(\log n)$ time respectively. 
Here $t= \min(\sqrt{\log \occ/\log \log \occ}, \log \log |P|)$, 
$\ndocs$ is the number of documents that contain $P$ 
and $\occ$ is the number of occurrences of $P$ in 
all documents.
\end{theorem}
Again, our query time does not depend on the pattern length, or this 
dependency is reduced.

\subsection{Compact Counting, Reporting and Document Reporting}
	\label{sec:compact}
	In this section, we show how our reporting and counting probems can be solved on {\em succinct} data structures \cite{Navarro:2007:CFI:1216370.1216372}. 

\subsubsection{Reporting and Counting.}
Our compact solution is based on compressed suffix
arrays~\cite{Grossi:2000:CSA:335305.335351}.  A compressed suffix
array for a text $T$ uses $|CSA|$ bits of space and enables us to
 retrieve the
position of the suffix of rank $r$, the rank of a suffix $T[i..]$, and
the character $T[i]$ in time $\Lookup(n)$.
Different trade-offs between space usage and query
time can be achieved (see~\cite{Navarro:2007:CFI:1216370.1216372} for
a survey). \no{; for instance, we can obtain $Lookup(n)=\log^{\varepsilon}n$  
and $|CSA|=O((1/\varepsilon)H_0n)$, where $H_0$ is the zero-order entropy 
and $\varepsilon$ is an arbitrary positive constant.
}
 \no{ There are several implementations of compressed suffix
  array with different trade-offs between space usage and query time
  (see~\cite{Navarro:2007:CFI:1216370.1216372} for a survey), here we
  use the one proposed
  in~\cite{Sadakane:2007:SDS:1224558.1224678}. The compressed suffix
  array of~\cite{Sadakane:2007:SDS:1224558.1224678} for $T$ can
  retrieve the position of the suffix of rank $r$, the rank of a
  suffix $T[i..]$, and the character $T[i]$ in time $\Lookup(n) =
  O(\log^{\epsilon} n)$, while using $|CSA| = O(n)$ bits.  }

Our data structure consists of a compressed generalized suffix array $CSA$ for 
$T_1,\ldots, T_m$  and compressed suffix arrays $CSA_i$ for each document 
$T_i$. In~\cite{Sadakane07} it was shown that using $O(n)$ 
extra bits, the length of the longest common
prefix of any two suffixes can be computed in $O(\Lookup(n))$ time. Besides, 
the ranks of any two suffixes $T_k[s..]$ and $T_{\ell}[p..]$ can be 
compared in $O(\Lookup(n))$ time: it suffices to compare $T_{\ell}[p+f]$ with 
$T_k[s+f]$ for $f=LCP(T_k[s..],T_{\ell}[p..])$.

Note that ranks of the suffixes of $T_\ell$ starting with $T_k[i..j]$ form an interval $[r_1, r_2]$. 
We use a binary search on the compressed suffix array of 
$T_\ell$ to find $r_1$ and $r_2$.   At each step of the binary search we 
compare a suffix of $T_{\ell}$ with $T_k[i..]$. Therefore $[r_1,r_2]$ 
can be found in $O(\Lookup(n)\cdot\log n)$ time. Obviously, the number of occurrences 
of $T_k[i..j]$ in $T_\ell$ is $r_2 - r_1$. To report the occurrences, we compute 
the suffixes of $T_\ell$ with ranks in interval $[r_1, r_2]$.

\begin{theorem}\label{theor:compact1}
All occurrences of $T_k[i..j]$ in $T_\ell$ can be counted in 
$O(\Lookup(n)\cdot\log{n})$ time and reported in
$O((\log n +\occ)\Lookup(n))$ time, where $\occ$ is the number of those. The
underlying indexing structure takes $2|CSA|+ O(n+m\log{\frac{n}{m}})$ bits of memory.
\end{theorem}

\subsubsection{Document Reporting}
Again, we use a binary search on the 
generalized suffix array to find the rank interval $[r_1,r_2]$ of suffixes 
that start with $T_k[i..j]$. This can be done in $O(\Lookup(n)\cdot\log n)$
 time.

In~\cite{Sadakane:2007:SDS:1224558.1224678}, it was shown how to report 
for any $1\le r_1\le r_2\le n$ all distinct documents $T_f$ such that at 
least one suffix of $T_f$ occurs at position $r$, $r_1\le r\le r_2$, 
of the generalized suffix array.  His construction uses 
$O(n+m\log\frac{n}{m})$ additional bits, and all relevant documents are reported 
in $O(\Lookup(n)\cdot\ndocs)$ time, where $\ndocs$ is the number of documents that 
contain $T_k[i..j]$. Summing up, we obtain the following result.
\begin{theorem}\label{theor:compact2}
All documents containing $T_k[i..j]$ can be reported in $O((\log n+ \ndocs)\Lookup(n))$ 
time, where $\ndocs$ is the number of those. The
underlying indexing structure takes $2|CSA|+O(n+m\log{\frac{n}{m}})$ bits
 of space.
\end{theorem}

\paragraph{\rm\bf Acknowledgments:} T.Starikovskaya has been supported by the
mobility grant funded by the French Ministry of Foreign Affairs
through the EGIDE agency and by a grant 10-01-93109-CNRS-a of the Russian Foundation 
for Basic Research. Part of this work has been done during
a one-month stay of Y.Nekrich at the Marne-la-Vall\'ee University
supported by the BEZOUT grant of the French government. 

\bibliographystyle{abbrv}

\bibliography{main}

\newpage
\section*{Appendix}
Here we prove Theorem~\ref{theor:wla}. We use the heavy path decomposition technique
  of~\cite{Amir:2007:DTS:1240233.1240242}.

\paragraph{Heavy Path Decomposition.}
\no{Our approach is based on  heavy path decomposition.}
 A path $\pi$ 
in $\cT$ is heavy if every node $u$ on $\pi$ has at most twice as many nodes 
in its subtree as its child $v$ on $\pi$.  A tree $\cT$ can be decomposed 
into paths using the following procedure:
we  find the longest heavy path $\pi_r$ that starts at the root of $\cT$
and remove all edges of $\pi_r$ from $\cT$.  
All remaining vertices of $\cT$ belong to a forest; we recursively 
repeat the same procedure in every tree of that forest.

We can represent the decomposition into heavy paths using a tree $\bT$. 
Each node $\bv_j$  in  $\bT$ corresponds to a heavy path 
$\pi_j$ in $\cT$. A node $\bv_j$ is a child of a node $\bv_i$ in 
$\bT$ if the head of $\pi_j$ (i.e., the highest node in $\pi_j$) 
is a child of some node $u\in \pi_i$.  
Some node in $\pi_i$ has at least twice as many descendants as each
node in $\pi_j$; hence, $\bT$ has height $O(\log n)$.

\paragraph{An $O(n\log n)$  Space Solution.}
Let $\bp_j$ denote a root-to-leaf path in $\bT$. 
For a node $\bv$ in $\bT$  let $weight(\bv)$ denote the weight 
of the  head of $\pi$, where 
$\pi$ is the heavy path represented by $\bv$ in $\bT$. 
We store a data structure $D(\bp_j)$ that contains the values of $weight(\bv)$ 
for all nodes $\bv\in \bp_j$.  
$D(\bp_j)$ contains $O(\log n)$ elements; hence, we can find the highest node 
$\bv\in \bp_j$ such that $weight(\bv)\ge q$ in $O(1)$ time. 
This can be achieved by storing  the weights of all nodes from $\bp_j$ in 
the q-heap\cite{FW94}. 

For every heavy path $\pi_j$ we store the weights of all 
nodes $u \in \pi_j$ in the data structure $E(\pi_j)$; using $E(\pi_j)$, we can 
find for any integer $q$ the  lightest node $u\in \pi_j$ 
such that $weight(u)\ge q$.
Using Theorem 1.5 in~\cite{AT07}, we can find the above defined 
node $u\in \pi_j$   in
 $O(\sqrt{\log n'/\log \log n'})$ time where $n'=\min(n_h,n_l)$,
$n_h=|\{\, v\in p_j\,|\,weight(v)> weight(u)\,\}|$, and 
$n_l= |\{\, v\in p_j\,|\,weight(v)< weight(u)\,\}|$. 
Moreover, we can also find the  node $u$ in $O(\log \log q)$ time;
we will describe the data structure in the full version of this paper. 
Thus $E(\pi_j)$ supports queries in $O(\min(\sqrt{\log n'/\log \log n'},\log\log q))$ time.


For each node $u\in \cT$ we store a pointer to 
the heavy path $\pi$ that contains $u$ and to the 
corresponding node $\bv\in \bT$.

A query $\anc(v,q)$ can be answered as follows. 
Let $\bv$ denote the node in $\bT$ that corresponds to the 
heavy path containing $v$. 
Let $\bp_j$ be an arbitrary root-to-leaf path in $\bT$ 
that also contains $\bv$. Using $D(\bp_j)$ we can find the 
highest node $\bu\in \bp_j$, such that $weight(\bu)\ge q$ in $O(1)$ time. 
Let $\pi_t$ denote the heavy path in $\cT$ that corresponds to 
the parent of $\bu$, and $\pi_s$ denote the path that corresponds 
to $\bu$.  If the weighted ancestor $\anc(v,q)$ is not the head of $\pi_s$, 
then  $\anc(v,q)$  belongs to the path  $\pi_t$. 
Using $E(\pi_t)$, we can find $u=\anc(v,q)$ in 
$O(\min(\sqrt{\log n'/\log \log n'}, \log \log q))$
time where $n'=\min(n_h,n_l)$,
$n_h=|\{\, v\in p_t\,|\,weight(v)> weight(u)\,\}|$, and 
$n_l= |\{\, v\in p_t\,|\,weight(v)< weight(u)\,\}|$. 

All data structures $E(\pi_i)$ use linear space.  Since there are $O(n)$ 
leaves  in $\bT$ and each path $\bp_i$ contains $O(\log n)$ nodes, 
all $D(\bp_i)$ use $O(n\log n)$ space.
\begin{lemma}
  \label{lemma:wla1}
There exists a $O(n\log n)$ space data structure that finds the weighted level ancestor $u$ in $O(\min(\sqrt{\log n'/\log \log n'}, \log \log q ))$
 time. 
\end{lemma}

\paragraph{An $O(n)$ Space Solution.}
We can reduce the space from $O(n\log n)$ to $O(n)$ using a micro-macro tree
 decomposition.  
Let $\cT_0$ be a tree induced by the nodes of $\cT$ that have at least 
$\log n/8$ descendants. The tree $\cT_0$ has at most $O(n/\log n)$ leaves. 
We construct the data structure described above for  $\cT_0$; since 
$\cT_0$ has $O(n/\log n)$ leaves, $\bT_0$ also has $O(n/\log n)$ leaves. 
Therefore all structures $D(\bp_j)$ use $O(n)$ words of space. 
All $E(\pi_i)$ also use $O(n)$ words of space.
If we remove all nodes of $\cT_0$ from $\cT$,  the remaining forest $\cF$
consists of $O(n)$ nodes. Every tree $\cT_i$, $i\ge 1$,
 in $\cF$ consists of 
$O(\log n)$ nodes. Nodes of $\cT_i$ are stored in a data structure 
that uses linear space
and answers weighted ancestor queries in $O(1)$ time. This data structure 
will be described later in this section.

Suppose that a weighted ancestor  $\anc(v,q)$ should be found. 
If $v\in \cT_0$, we answer the query using the data structure for $\cT_0$. 
If $v$ belongs to some $\cT_i$ for $i\ge 1$,
 we check the weight $w_r$ of $root(\cT_i)$. 
If $w_r\le q$, we search for $\anc(v,q)$ in $\cT_i$.
Otherwise we identify the parent $v_1$ of $root(\cT_i)$ and 
find $\anc(v_1,q)$ in $\cT_0$. If $\anc(v_i,q)$ in $\cT_0$  is undefined, 
then $\anc(v,q)=root(\cT_i)$.

\paragraph{A Data Structure for a Small Tree.}
It remains to describe the data structure for a tree $\cT_i$, $i\ge 1$. 
Since $\cT_i$ contains a small number of nodes, we can answer weighted 
level ancestor queries on $\cT_i$ using a look-up table $V$.
$V$ contains information about any tree with up to 
$\log n/8$ nodes, such that node weights are positive 
integers bounded by $\log n/8$.
For any such tree $\tT$, for any node $v$ of $\tT$, and for any integer 
$q\in [1,\log n/8]$, we store the pointer to $\wla(v,q)$ in $\tT$. 
There are $O(2^{\log n/4})$ different trees $\tT$ (see e.g.,~\cite{BenderF04}
 for a simple proof); 
for any $\tT$, we can assign weights to nodes in less than 
$(\log n/8)!$ ways. 
For any weighted tree $\tT$ 
there are at most 
$(\log n)^2/64$ different pairs $v$, $q$. Hence, the table $V$ contains 
$O(2^{\log n/4}(\log n)^2(\log n/8)!)=o(n)$ 
entries. 
We need only one look-up table $V$ for all mini-trees $\cT_i$.

We can now answer a weighted level ancestor query on $\cT_i$ using 
reduction to rank space.  The \emph{rank} of a node $u$ in a tree 
$\cT$ is defined as $\rrank(u,\cT)=|\{\,v\in \cT\,|\, weight(v)\le weight(u)\,\}|$. 
The successor of an integer $q$ in a tree $\cT$ is the lightest node $u\in \cT$
such that $weight(u)\ge q$. 
The rank $\rrank(q,\cT)$ 
of an integer $q$  is defined as the rank of its successor. 
\no{equals to $|\{\,v\in \cT\,|\, weight(v)\le q \,\}|$. }
Let $\rrank(\cT)$ denote the tree $\cT$ in which the weight of every node 
is replaced with its rank.  
The weight of a node $u\in \cT$ is not smaller than $q$ if an only if 
$\rrank(u,\cT)\ge \rrank(q,\cT)$. Therefore we can find $\wla(v,q)$ in 
some $\cT_i$ as follows. For every $\cT_i$ we store a pointer to 
$\tT_i=\rrank(\cT_i)$. 
Given a query $\wla(v,q)$, we find $\rrank(q,\cT_i)$ in $O(1)$ time
 using a q-heap~\cite{FW94}. 
Let $v'$ be the node in $\tT_i$ that corresponds to the node $v$.
We find $u'=\wla(v',\rrank(q,\cT_i))$ in $\tT_i$ using the table $V$. 
Then the node $u$ in $\cT_i$ that corresponds to $u'$ is the weighted level 
ancestor of $v$.

\end{document}